\documentclass[%
 aip,
 amsmath,amssymb,
 reprint,%
]{revtex4-1}

\usepackage{graphicx}
\usepackage{dcolumn}
\usepackage{bm}

\usepackage[utf8]{inputenc}
\usepackage[T1]{fontenc}
\usepackage{mathptmx}
\usepackage{etoolbox}
\usepackage{ulem}
\usepackage{xcolor}
\usepackage{physics}
\makeatletter
\def\@email#1#2{%
 \endgroup
 \patchcmd{\titleblock@produce}
  {\frontmatter@RRAPformat}
  {\frontmatter@RRAPformat{\produce@RRAP{*#1\href{mailto:#2}{#2}}}\frontmatter@RRAPformat}
  {}{}
}%
\makeatother
\begin{document}

\preprint{AIP/123-QED}

\title{Investigation of fluorescence versus transmission readout for three-photon Rydberg excitation used in electrometry}

\author{Nikunjkumar Prajapati}
\affiliation{ National Institute of Standards and Technology, Boulder, Colorado 80305, USA}
 \email{nikunjkumar.prajapati@nist.gov}
\author{Samuel Berweger}
\affiliation{ National Institute of Standards and Technology, Boulder, Colorado 80305, USA}
\author{Andrew P. Rotunno}
\affiliation{ National Institute of Standards and Technology, Boulder, Colorado 80305, USA}
\author{Alexandra B. Artusio-Glimpse}
\affiliation{ National Institute of Standards and Technology, Boulder, Colorado 80305, USA}
\author{Noah~Schlossberger}
\affiliation{Associate of the National Institute of Standards and Technology, Boulder, Colorado 80305, USA}
 \affiliation{Department of Physics, University of Colorado, Boulder, Colorado 80302, USA}
\author{Dangka Shylla}
 \affiliation{Department of Physics, University of Colorado, Boulder, Colorado 80302, USA}
\affiliation{Associate of the National Institute of Standards and Technology, Boulder, Colorado 80305, USA}
\author{William J. Watterson}
 \affiliation{Department of Physics, University of Colorado, Boulder, Colorado 80302, USA}
\affiliation{Associate of the National Institute of Standards and Technology, Boulder, Colorado 80305, USA}
\author{Matthew T. Simons}
\affiliation{ National Institute of Standards and Technology, Boulder, Colorado 80305, USA}
\author{David LaMantia}
\affiliation{ National Institute of Standards and Technology, Gaithersburg, MD 20899 USA}
\author{Eric B. Norrgard}
\affiliation{ National Institute of Standards and Technology, Gaithersburg, MD 20899 USA}
\author{Stephen P. Eckel}
\affiliation{ National Institute of Standards and Technology, Gaithersburg, MD 20899 USA}
\author{Christopher L. Holloway}
\affiliation{ National Institute of Standards and Technology, Boulder, Colorado 80305, USA}
\date{\today}

\begin{abstract}
We present a three-photon based fluorescence readout method where the strength of the fluorescence scales with the strength of the radio-frequency (RF) field being applied. 
We compare this method to conventional three-photon electromagnetically-induced transparency (EIT) and electromagnetically-induced absorption (EIA). 
Our demonstrated EIA/EIT sensitivity in the collinear three-photon Cesium system is the best reported to date at roughly 30 $\mu Vm^{-1}Hz^{-1/2}$.
The fluorescence is nearly 4 fold better in senstivity compared to EIA/EIT readout. 
\end{abstract}

\maketitle

\section{Introduction}
Rydberg atoms, which refer to atoms excited to a high principal quantum number, have emerged as a highly advantageous resource in various domains~\cite{gallagher_book}. Rydberg atoms possess a significant dipole moment and enable measurements linked to the International System of Units (SI)~\cite{gor1,first_ryd_elet_shaff,6910267,si_trace}. The exceptional sensitivity and traceability offered by alkali Rydberg atoms have rendered them an invaluable asset in numerous diverse fields.

Rydberg atoms are used to measure AC/DC electric fields~\cite{10.1116/5.0090892,doi:10.1116/5.0097746,PhysRevApplied.18.024001}, quantum computation and simulation~\cite{quant_info_ryd,entanglement_ryd}, single-photon detection and generation~\cite{ryd_single_photon,PhysRevLett.117.223001}, and quantum storage~\cite{quant_info_ryd}. Their use as an electric field sensor alone has a long history of development and applications\cite{9748947}. These applications include using Rydberg atoms as a measurement standard~\cite{gor1,first_ryd_elet_shaff,6910267,si_trace,Norrgard_2021,holl1,7812705}, a power standard~\cite{pow_stand}, Rydberg thermometry~\cite{Norrgard_2021}, receiving amplitude-, frequency-, and phase-modulated signals~\cite{doi:10.1063/1.5028357,8778739,doi:10.1063/1.5088821,9054945,doi:10.1116/5.0098057}, for beam-forming~\cite{synthetic_app_ryd}, use as a broadband spectrum analyzer~\cite{waveguide_SA}, and many others~\cite{terehertz_imaging,aoa,Kumar2017,doi:10.1063/1.5099036}. In the majority of these applications, Rydberg atoms are generated and probed via two-photon electromagnetically-induced transparency (EIT) schemes. However, three-photon excitation schemes are emerging\cite{PhysRevA.100.063427,10.1117/12.2586718,Carr:12,10.1117/12.2309386,10.1063/5.0147827}.

The conventional ladder-EIT approach benefits from experimental simplicity and well-established implementations~\cite{Tanasittikosol_2011,Sedlacek_2013,Norrgard_2021,first_ryd_elet_shaff}, but new detection methods have emerged recently.
These involve more complex EIT schemes such as a V-configuration~\cite{Wang:17}, as well as fluorescence~\cite{10.1116/5.0090892}, and six-wave mixing~\cite{6wavemix}.
Although the latter is more complex and can require a single-photon detector for readout, it can achieve exceptional sensitivity by eliminating the large signal background in EIT.
Similarly, monitoring the Rydberg state fluorescence can be used for field sensing with little or no signal background, though it has not been extensively explored~\cite{gallagher_book}. 
There are many decay pathways available that could be used for fluorescence to measure the Rydberg state, and the choice of these is complex.
Of the available transitions in cesium ($^{133}$Cs), the 510~nm decay to the 6P$_{3/2}$ state is particularly attractive due the branching ratios, but it cannot be readily used in two-photon implementations because of its use in the excitation ladder and the fact that it is no longer dipole allowed after a transition between Rydberg states has occurred.

\begin{figure}
    \centering
    \includegraphics[width = \columnwidth]{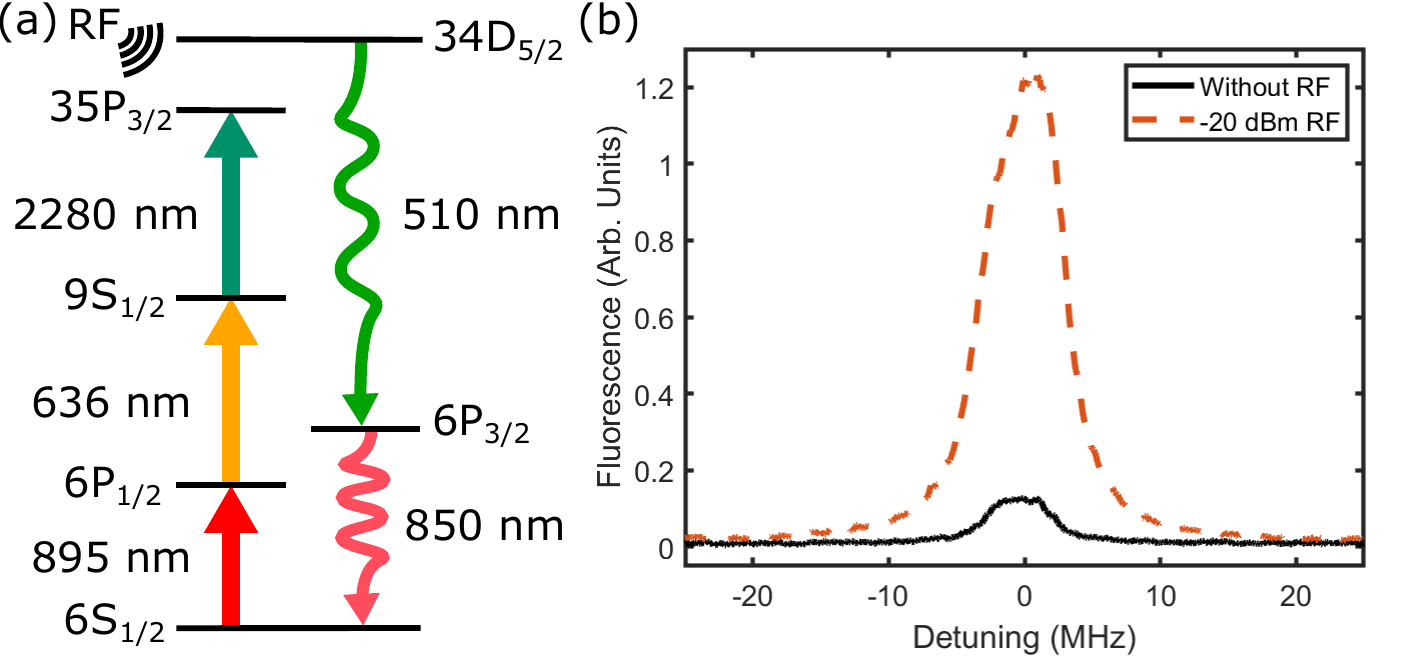}
    \caption{(a) $^{133}$Cs level diagram showing three-photon excitation to 35P$_{3/2}$ state and subsequent fluorescence. (b) Sample spectra of fluorescence at 510 nm with (red-dashed) and without RF applied (black-solid) as the coupling laser (2280 nm) frequency is scanned.}
    \label{fig:level_diagram}
\end{figure}

In this manuscript, we compare the measurement of three-photon electromagnetically-induced absorption (EIA)/EIT detection and fluorescence-based detection. Fig.~\ref{fig:level_diagram} (a) shows the excitation path we use to observe three-photon EIA/EIT, where we read the D2 probe line transmission in $^{133}$Cs. 
We only observe the generation of 510 nm photons from the Rydberg D state decay. 
In prior fluorescence measurements, the readout was from the decay of the Rydberg state being excited, which will have a background photon count irrespective of the radio-frequency (RF)~\cite{10.1116/5.0090892}. 
However, our measurement requires an applied RF field to generate the 510 nm photons, shown by Fig.~\ref{fig:level_diagram} (b). 
This type of measurement can be a useful tool for characterizing interesting phenomena like black-body radiation-based state transfer or state-changing collisions between Rydberg atoms. We will discuss this more in section~\ref{sec:resid_flo}.

\begin{figure*}[ht]
    \centering
    \includegraphics[width = .8\textwidth]{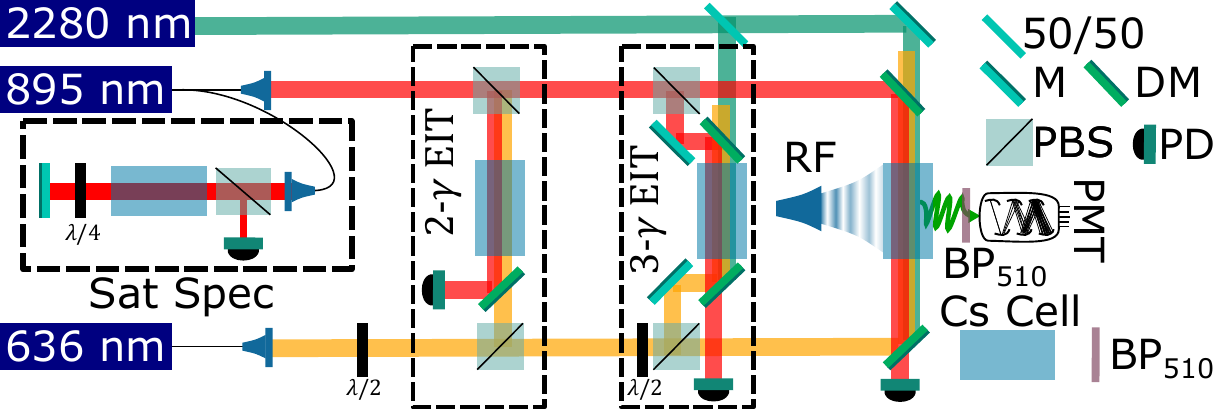}
    \caption{Experimental apparatus showing the three lasers. 
    The saturation spectroscopy (Sat Spec) is used for feedback to lock the 895 nm laser. 
    The 2-photon EIT (2-$\gamma$ EIT) is used for feed back to lock the 636 nm laser.
    The three-photon EIT (3-$\gamma$ EIT) is used to stabilize the 2280 nm laser.
    The figures list shows a 50/50 beam splitter (50/50), mirror (M), dichroic mirror (DM), polarizing beam-splitter (PBS), photo-detector (PD), photo-multiplier tube (PMT),  zero-order quarter wave-plate ($\lambda$/4), zero-order half wave-plate ($\lambda$/2), and 510 nm bandpass filter (BP$_{510}$).}
    \label{fig:experimental_apparatus}
\end{figure*}

In the next section, we discuss the experimental apparatus followed by a comparison of spectra generated for EIA/EIT and fluorescence. Then, we map out how the signal amplitude changes for varying probe powers and applied RF fields. We then compare the sensitivity map for the two methods and demonstrate that the fluorescence operates better for the conditions we utilized here. Finally, we present intriguing findings that may hint at the presence of additional photons from either Rydberg state collisions or black-body radiation. 

\section{Experimental Apparatus}
We generate the EIA/EIT and fluorescence signals utilizing a three-photon excitation with 895 nm, 636 nm, and 2280 nm lasers. The 895 nm laser and 636 nm optical fields are generated by external-cavity diode lasers (ECDLs). The 2280 nm optical field was generated by an amplified 1068 nm ECDL passed through an optical parametric oscillator (OPO) crystal that generates a 2012 nm signal and a 2280 nm idler field. 

Figure.~\ref{fig:experimental_apparatus} depicts the laser locking and experimental setup. We lock the 895 nm probe optical field to the $\ket{6S_{1/2}, F=3} \rightarrow \ket{6P_{1/2}, F=4}$ transition using saturated spectroscopy signal. We lock the 636 nm dressing optical field using the 2-photon EIT generated by the 895 nm and 636 nm field.
The 636 nm laser was tuned to the $\ket{6P_{1/2}, F=4} \rightarrow \ket{9S_{1/2}, F=3}$ transition. For the 2280 nm laser, we have a two-step lock. The 1068 nm fundamental is line narrowed by locking to an ultra-low expansion (ULE) cavity under vacuum, not shown. The idler (2280 nm coupling optical field) was stabilized to the three-photon EIA/EIT signal feeding back to the OPO. This method allowed us to narrow the linewidth of the 2280 nm optical field while locking to the EIT feature.

In the cell used for the experiment, we read out the EIA/EIT using a balanced photo-detector and the fluorescence using a photo multiplier tube (PMT) (Thorlabs PMM01-PMT \cite{NISTDisclaimer}). The PMT has three 510 nm band-pass filters with 10 nm full width at half max placed on it to isolate Rydberg decays from the S, D, and F states. The three lasers in the ladder are co-linear and counter-propagating so that there is minimal Doppler residual, as demonstrated in Shaffer et. al.\cite{10.1117/12.2309386} and Prajapati et. al.\cite{10.1063/5.0147827}. The 895 nm, 636 nm, and 2280 nm optical fields had 900 $\mu$m, 880 $\mu$m, and 850 mm 1/$e^2$ beam radius. Ideally, the probe would be the smallest of the three, but we assume that their differences are small enough to not have a strong effect. For the results presented, the Rabi frequency of the 636 nm and 2280 nm lasers were 2$\pi\cdot$7 MHz and 2$\pi\cdot$10 MHz, respectively. This is for optical powers of 10 mW for the 636 nm laser at the cell and 1000 mW of power for the 2280 nm laser. 

\section{EIA/EIT vs Fluorescence Readout}
\begin{figure}[ht]
    \centering
    \def\widthScale{0.5}
    \includegraphics[width = \widthScale\columnwidth]{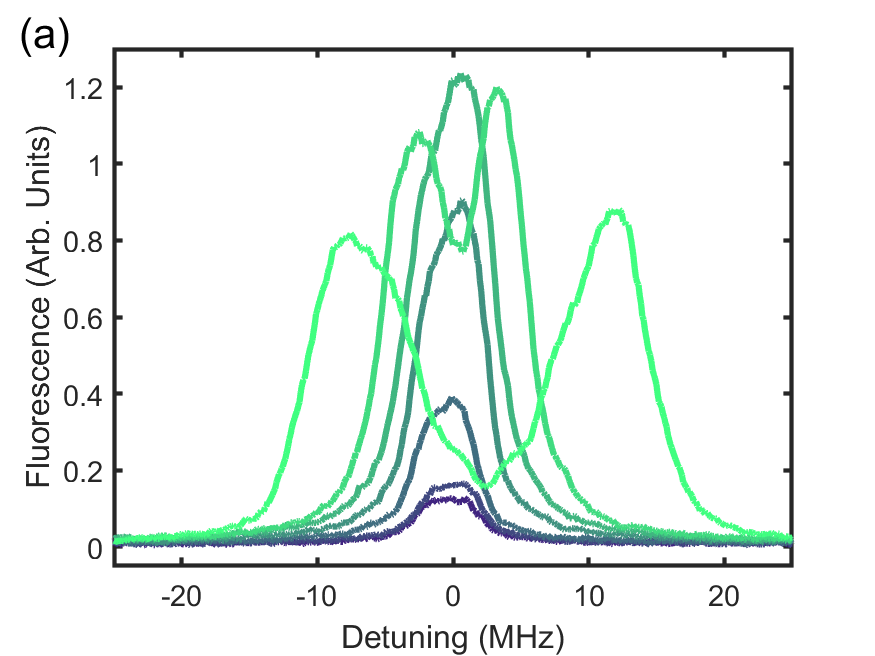}\includegraphics[width = \widthScale\columnwidth]{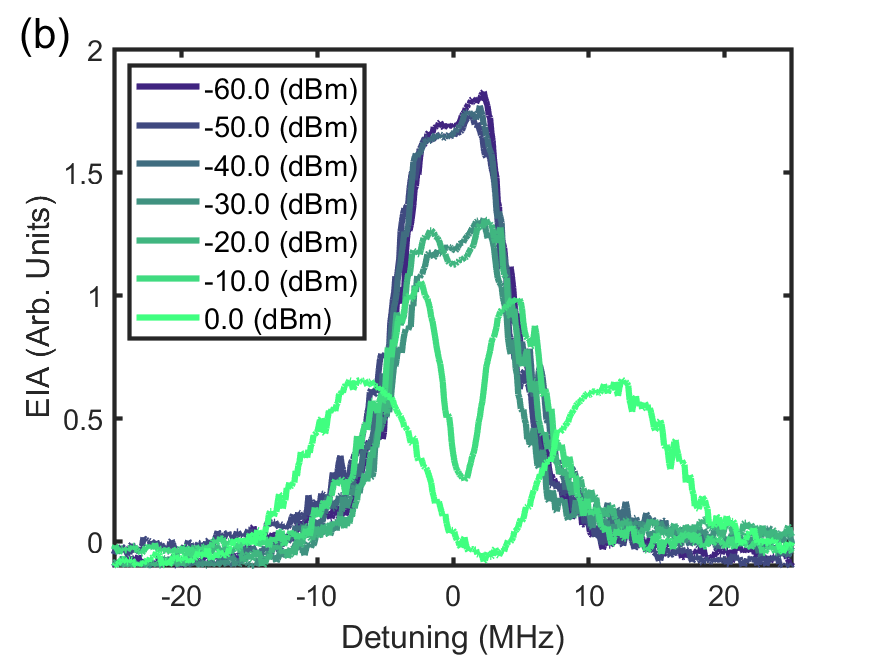}
    \caption{(a) Fluorescence spectrum as a function of the coupling laser detuning. (b) EIA spectrum as a function of the coupling laser detuning. The different colored traces are for different applied RF powers. The legend given in (b) is for both figures (a) and (b).}
    \label{fig:sample_spectra}
\end{figure}

Figure.~\ref{fig:sample_spectra} (a) shows the detected 510 nm fluorescence as a function of coupling laser detuning for various applied RF powers. The lowest applied power of -60 dBm RF is equivalent to the RF field being off since we run into background fluorescence. We clearly increased fluorescence with applied powers as low as -50 dBm. As the RF power is increased, we can see that the fluorescence peak begins to increase until it reaches a maximum and then splits. When we compare this to the case of the three-photon EIA (shown by Fig.~\ref{fig:sample_spectra} (b)), we see that there is no response until an RF power of more than -30 dBm is applied. This already shows that the fluorescence responds to weaker RF fields than the EIT. For the case of fluorescence readout, there is a signal remnant even when no RF is applied, shown by trace for -60 dBm RF in Fig.~\ref{fig:sample_spectra} (a). This limits how weak of a fluorescence we can observe due to an RF field. We discuss this feature in more detail in Sec.~\ref{sec:resid_flo}.

\begin{figure}[ht]
    \centering
    \def\widthScale{1}
    \includegraphics[width = \widthScale\columnwidth]{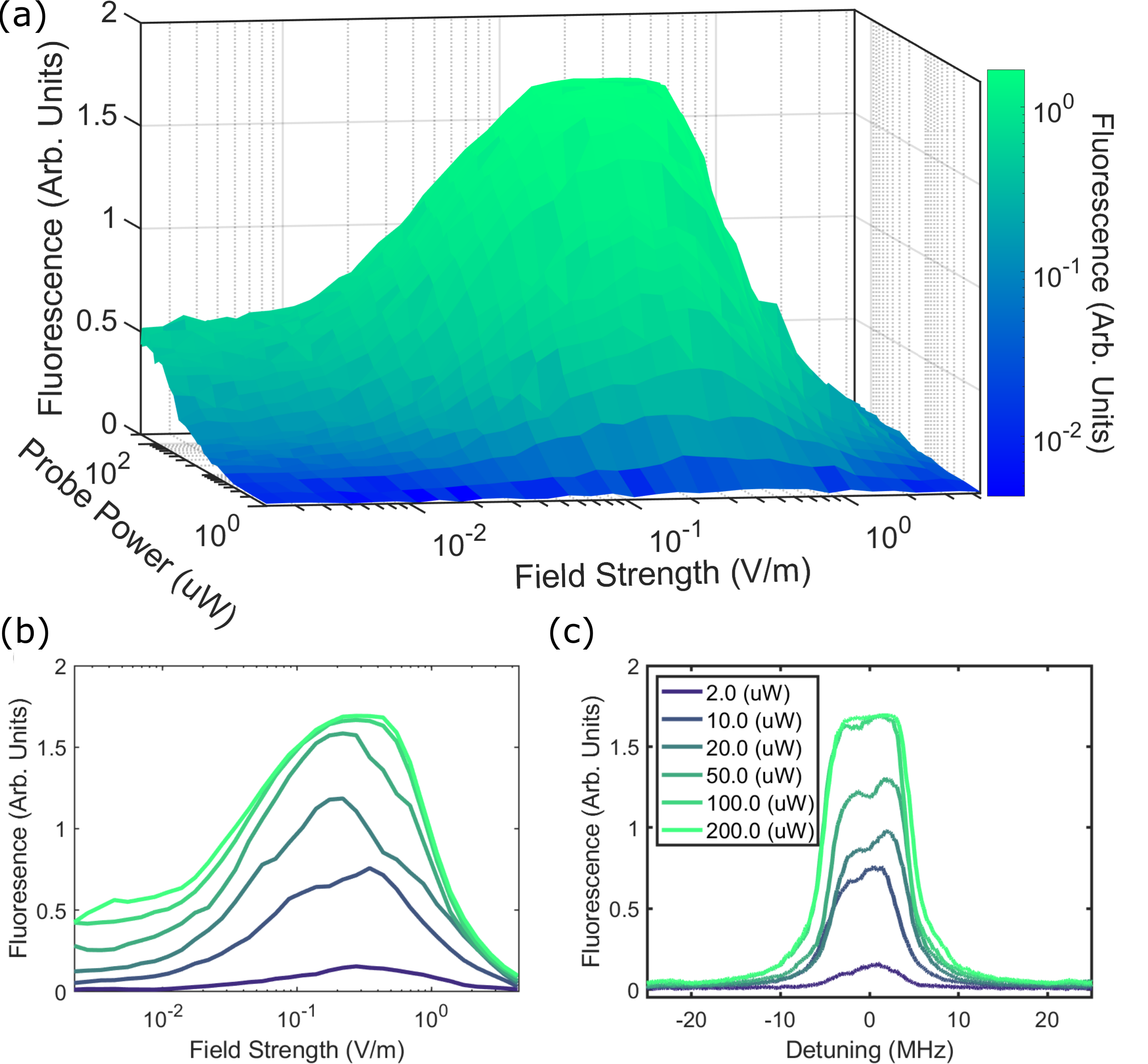}
    \caption{(a) 3D surface showing the dependence of fluorescence with lasers locked to atomic resonances described in text for different RF field strengths and probe optical powers. (b) Resonant fluorescence value extracted from (a). Each trace shows the dependence on electric field strength and the different traces are for different probe laser powers. (c) Fluorescence resonance as coupling laser is scanned. The different traces are for different probe laser powers. (b) and (c) share the same legend.}
    \label{fig:FLO_waterfall}
\end{figure}

\begin{figure}[ht]
    \centering
    \def\widthScale{.9}
    \includegraphics[width = \widthScale\columnwidth]{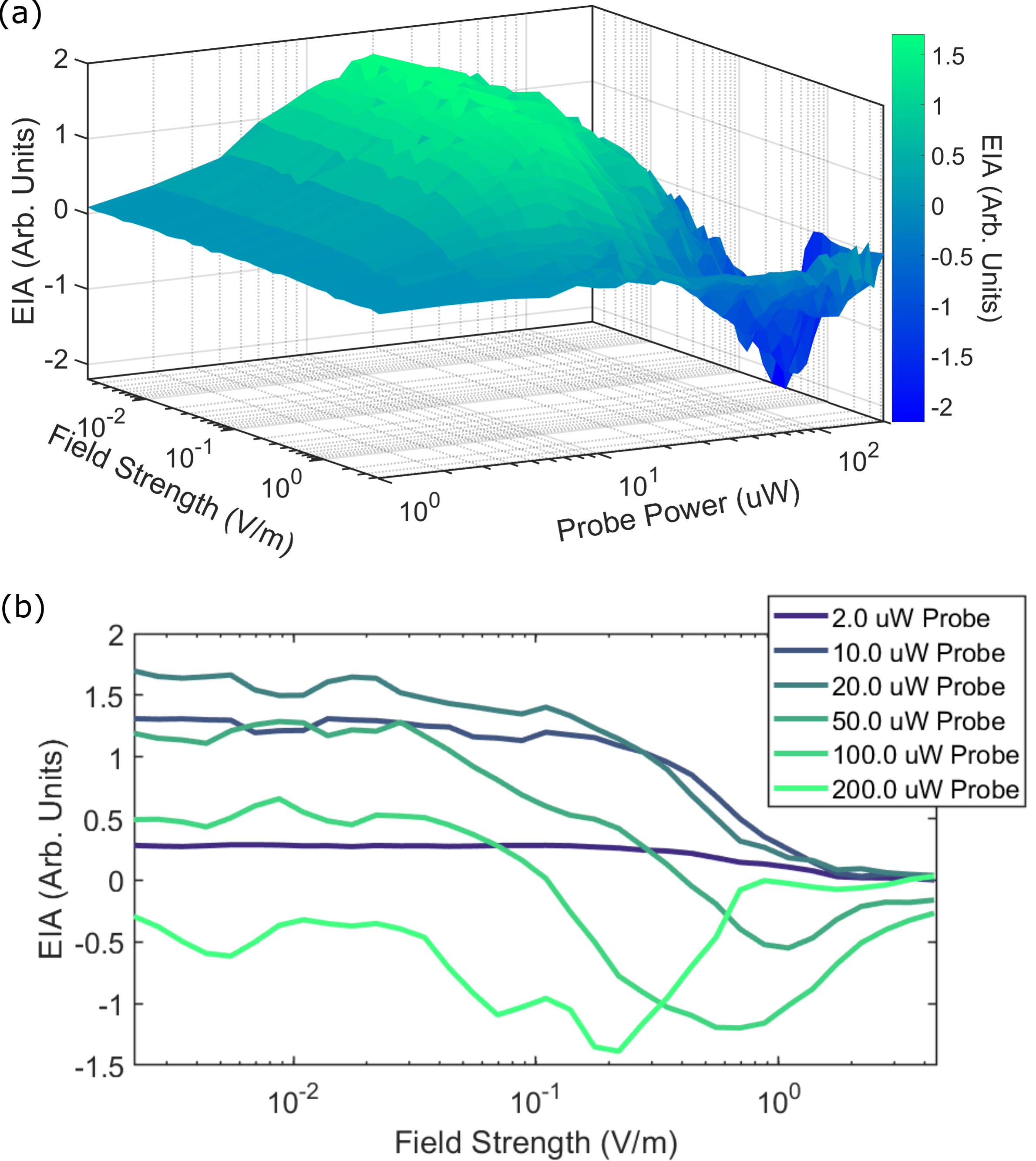}
    \caption{(a) 3D surface showing the dependence of EIA with lasers locked to atomic resonances described in text for different RF field strengths and probe optical powers. (b) Resonant EIA value extracted from (a). Each trace shows the dependence on electric field strength and the different traces are for different probe laser powers, shown by legend.}
    \label{fig:EIT_waterfall}
\end{figure}

We map out the resonant fluorescence and EIA values for various values of probe optical power and RF field strength, shown by Fig.~\ref{fig:FLO_waterfall} and Fig.~\ref{fig:EIT_waterfall}, respectively. We see an increase in fluorescence amplitude in response to the field strength around 1 mV/m. While, for the case of EIA, we don't see the EIA amplitude respond until over 20 mV/m. In Fig.~\ref{fig:FLO_waterfall} (a), (b), and  (c), we observe a fluorescence that scales withe probe power in a region insensitive to the RF field. Another interesting response to the field strength for large probe powers is a saturation effect that emerges. Fig.~\ref{fig:FLO_waterfall} (a) shows that there is a region where the fluorescence flattops for field strengths of 10 mV/m to 100 mV/m at the highest probe power. This is better seen in Fig.~\ref{fig:FLO_waterfall} (b) and Fig.~\ref{fig:FLO_waterfall} (c). After incorporating a neutral density filter in front of the PMT, we continue to observe this flattop characteristic despite its initial appearance of being electronic in nature.

Several intriguing features are observed in the EIA/EIT spectrum, which are evident in the theoretical analysis. Among these, a notable characteristic is the transition from EIT to EIA as we increase the probe power. In the theory, the crossover point occurs when the probe Rabi rate is 3/5 of the coupling Rabi rate~\cite{10.1063/5.0147827}. This observation aligns with the data we have obtained. Specifically, our coupling Rabi rate measures approximately 10 MHz, whereas the probe Rabi rate at the zero EIA value is approximately 6.2 MHz for a very weak RF field.  Another characteristic matching the theory is the cross-over from EIA to EIT produced from an applied RF field. This trend can be seen more clearly in Fig.~\ref{fig:EIT_waterfall} (b). For increasing probe laser powers, we see that the peak begins as an EIT feature and then moves to EIA, shown by trace for 100 uW probe power in Fig.~\ref{fig:EIT_waterfall} (b). 

\begin{figure}
    \centering
    \includegraphics[width = \columnwidth]{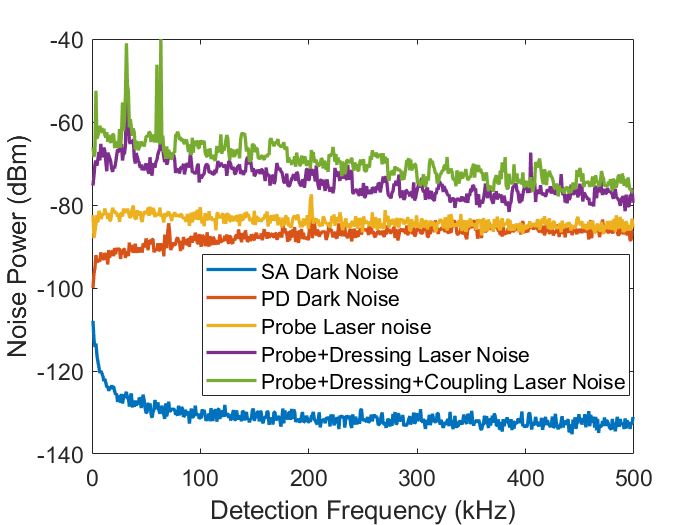}
    \caption{Noise present in system from different sources as labeled. In order, the different noises are Spectrum analyzer (SA) dark noise, photo-detector (PD) dark noise, probe laser noise locked to atomic resonance, probe laser noise with dressing laser locked to resonance, and probe laser noise with dressing and coupling lasers locked to respective atomic resonances.}
    \label{fig:noise}
\end{figure}

Regrettably, stronger oscillations emerge in the transmission when the power of the probe laser is increased. This phenomenon is associated with two primary factors. Firstly, there is an escalation in the noise of the probe laser. Secondly, there is an augmentation in the two-photon coupling between the dressing and probe laser fields. We have observed that laser noise on the dressing optical field will transfer through the atoms onto the transmission of the probe laser, shown in Fig.~\ref{fig:noise}. Despite the heightened fluctuations, the EIA/EIT measurement has increased electric field sensitivity when utilizing higher probe laser powers.

\section{Sensitivity of EIA/EIT vs Fluorescence}
\label{SSFL}

We measure the sensitivity by performing an atom heterodyne measurement, following the methodology outlined in the publication by Gordon et. al.\cite{gor3} We apply a local oscillator (LO) RF field that beats with the signal RF field and use a spectrum analyzer to determine the signal-to-noise (SNR) of the beat note for a specific set of parameters under examination. The SNR is then utilized to determine the minimum detectable field for a 1-second measurement time, in other words, sensitivity. For instance, in the case where the SNR is determined to be 25 dB for an applied RF signal power of $P_{\rm{sig}} = -60$ dBm with an SA resolution bandwidth (f$_{RBW}$) of 10 Hz, the resulting sensitivity $S$ or noise equivalent field for the measurement can be calculated by
\begin{equation}
    S = \sqrt{10^{(P_{\rm{sig}}-SNR)/10}\cdot f_{RBW}}\cdot C_{\rm{cal}} = 38 \frac{uV}{m\sqrt{Hz}},
    \label{eq:cal}
\end{equation}
where $C_{\rm{cal}}$ is the calibration factor obtained from fitting to the AT splitting plotted against the applied RF signal power. We utilized the AT splitting to calibrate the power of the applied RF signal generator to field strength at the cell. This calibration process is described in Holloway et. al.~\cite{si_trace} and involves scanning the coupling laser to measure the AT splitting, as shown in Fig.~\ref{fig:sample_spectra} (a). Instead of using the conventional approach of employing EIT for the measurement, we opted to utilize fluorescence. We calibrated the coupling laser scan by sinusoidal modulating the coupling laser current at a frequency of 5 MHz. This modulation generated side bands on the signal from the ULE cavity which was used to scale the time trace from the oscilloscope. We used the data from  Fig.~\ref{fig:sample_spectra} (a) for calibrating the electric field. 

\begin{figure}
    \centering
    \includegraphics[width = \columnwidth]{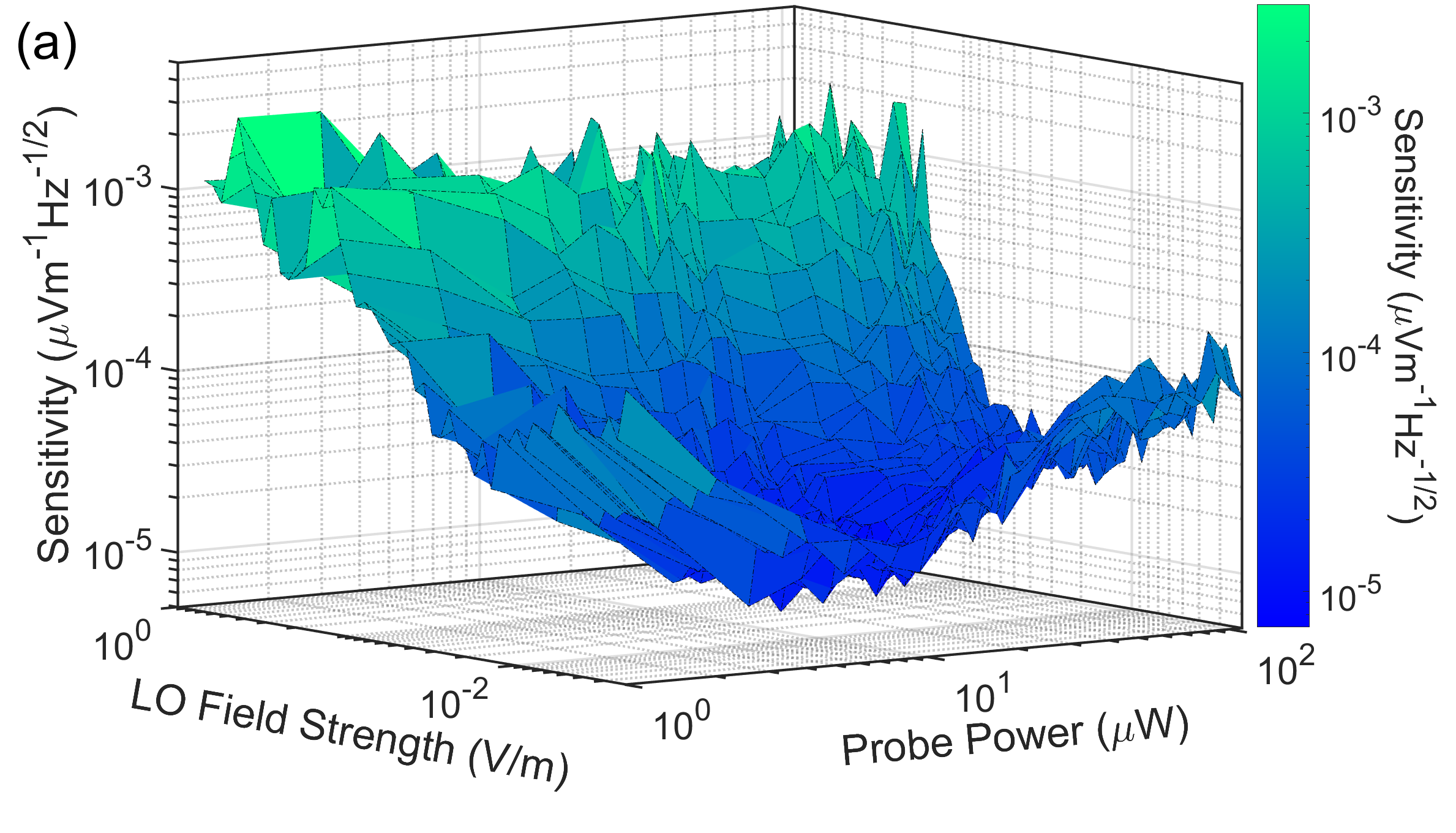}
    \includegraphics[width = \columnwidth]{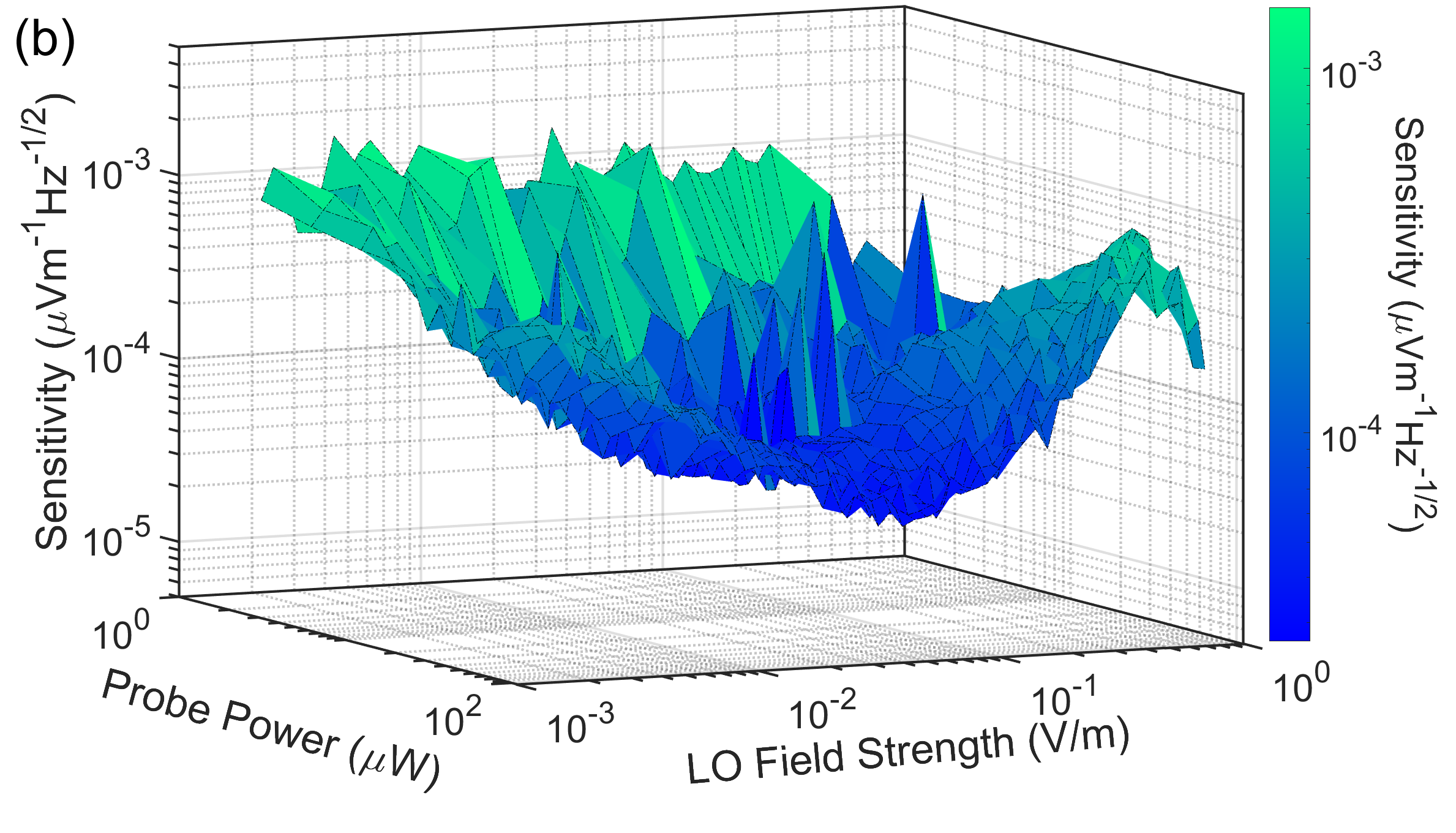}
    \caption{(a) Sensitivity of fluorescence detection for different probe laser powers and RF LO field strengths. (b) Sensitivity of EIA/EIT detection for different probe laser powers and RF LO field strengths.}
    \label{fig:sensitivity}
\end{figure}

We mapped the electric field sensitivity of the EIA/EIT and fluorescence for various probe laser powers and LO RF powers. For these measurements, the signal RF power was fixed to -60 dBm (70 $\rm{\mu V/m}$). We used two spectrum analyzers to simultaneously measure the SNR of the EIA/EIT signal and the fluorescence signal. We then used Eq.~\ref{eq:cal} to calculate the sensitivity using the measured SNR. For the case of fluorescence, the best sensitivity we measured was 8~$\mu V m^{-1} Hz^{-1/2}$. For the case of EIA/EIT, the best sensitivity we measured was 30~$\mu V m^{-1} Hz^{-1/2}$. The fluorescence RF electrometry measurements demonstrated here not only outperform those of EIA/EIT, but are also the best sensitivity shown to-date for the three-photon system in cesium. Previous measurements in two-photon EIT have demonstrated 3 $\mu m^{-1}Hz^{-1/2}$ in sensitivity, but this utilized a repump field~\cite{Repump_paper}. We should be able to improve our sensitivity by using a repumping laser~\cite{Repump_paper}.

Another noteworthy point here is the difference in optimal RF LO power. The optimal RF LO power for the fluorescence was roughly 5 mV/m while for the optimal RF LO power for the EIA/EIT was roughly 40 mV/m. This is an important point for applications where little to no radiation is wanted and power consumption is a problem. Reducing this by nearly an order of magnitude in field is a massive benefit. 

\section{Bandwidth of EIA/EIT VS Fluorescence}
\begin{figure*}[ht]
    \centering
    \includegraphics[width=\textwidth]{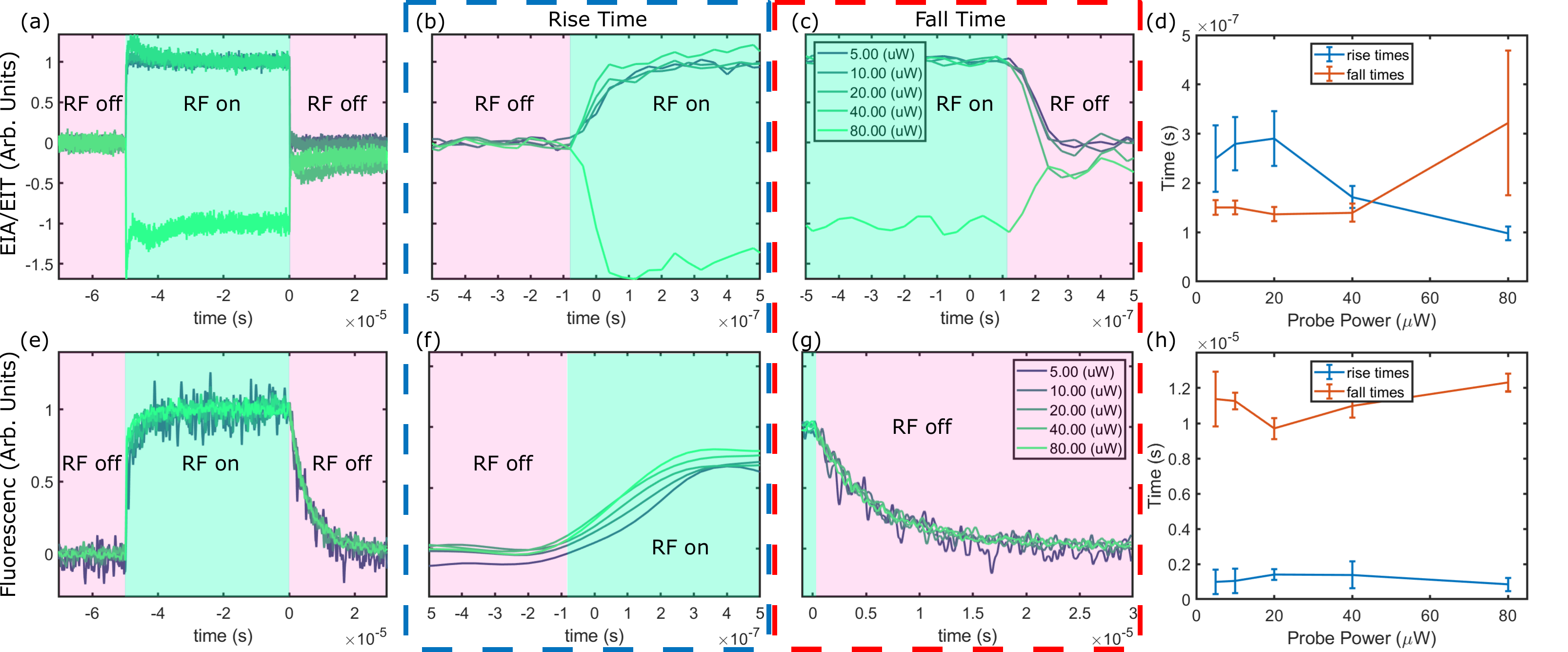}
    \caption{(a,e) EIA/EIT and fluorescence response to RF field with square-wave modulation as labeled. (b,f) Rise time: effect on EIA/EIT and fluorescence of the applied RF field. (c,g) Fall time: effect on EIA/EIT and fluorescence of the applied RF field. (d,h) Extracted rise and fall times from (b,c,f,g) as labeled. As a clarifying note, top row are measurments of EIA/EIT response while bottom row is fluorescence response. The rise and fall times are the mean and the errorbars are the standard deviation of 10 traces taken for one power. The legend in Fig (c,g) are shared for for figures (a-c) and (e-g). } 
    \label{fig:bandwidth}
\end{figure*}

We characterize the bandwidth of the two systems by assessing the rise and fall times of the signal in response to the RF. The bandwidth of Rydberg atom systems has been of interest in various different implementations and the investigation of bandwidth limitations is still ongoing. Different groups discuss the population and decoherence of the dark state being a limiting aspect~\cite{doi:10.1116/5.0098057,Hu_Jiao_He_Zhang_Zhang_Zhao_Jia_2023}, others argue the limits in the power of the optical fields that induce Rabi flopping. We discuss potential causes for limited bandwidth here and present the bandwidth of the EIA/EIT and fluorescence measurements.

Figure~\ref{fig:bandwidth} shows the rise and falls times for the EIA/EIT and fluorescence measurements. We immediately notice that the response the EIA/EIT is much faster than that of the fluorescence. We define the rise time as the time it takes for the EIA/EIT and fluorescence to respond to the RF field turning on, shown by red dashed box in Fig.~\ref{fig:bandwidth} (b) and (f). The fall time is the time it takes for EIA/EIT and fluorescence to respond to the RF field turning off, shown by blue dashed box in Fig.~\ref{fig:bandwidth} (c) and (g). 

The extracted times are the 90/10 fits of the rise and fall traces, shown in Fig.~\ref{fig:bandwidth} (d) and (h). We can see that the EIA/EIT has a rise and fall time of roughly 200 ns. For the optimal power of $\approx$50$~\mu W$, the rise and fall time are roughly 150 ns. We can determine the bandwidth from the rise and fall times using BW = 0.35/$\tau$ ~\cite{Bogatin_2020}, where $BW$ is bandwidth and $\tau$ is the rise or fall time. The bandwidth from this is then $\approx$3 MHz. In EIA/EIT, the rise will be governed by the time it takes for EIA/EIT to vanish, in other words the decay time of the dark state. The dark state decay is proportional to the effects of collisions, radiative decay, and black-body radiation (BBR) induced state transfer. We calculate the total decay rate out of the 34 D Rydberg state to be $\approx$10*2$\pi$ kHz by utilizing the ARC Rydberg calculator~\cite{vsibalic2017arc,NISTDisclaimer}. Then by using the collisional cross section of the 34 D Rydberg state and the mean free path of a room temperature Cesium atom, we find the collision rate of the Rydberg atoms with other Rydberg atoms to be $\approx$10*2$\pi$ kHz. These result in a total decay time of 50 $\mu s$. However, this is not the time observed, and the response is much faster, shown by Fig.~\ref{fig:bandwidth} (d). This is likely from the fact that the three-photon system does not have a dark state, unlike conventional two-photon Rydberg excitation's. The fall time is simply governed by the Rabi rate of the EIA/EIT process. 

The fluorescence on the other hand has very different rise and fall times. The rise time is governed by the Rabi rate of the RF field driving the population transfer to the 34D Rydberg state. This Rabi rate was on the order of several MHz. The decay rate is set by the decay of the 34D Rydberg state. Again, this decay time is on the order of 50 $\mu s$ and is set by the collisions, radiative decay, and black-body radiation. We discuss these factors in more detail in the next section. From Fig.~\ref{fig:bandwidth} (h) we see that the fall time is roughly 12 $\mu s$. This means that the residual fluorescence is likely from the transfer of population from BBR and collisions effects.

\section{Residual Fluorescence}
\label{sec:resid_flo}
\begin{figure}[ht]
    \centering
    \includegraphics[width=\columnwidth]{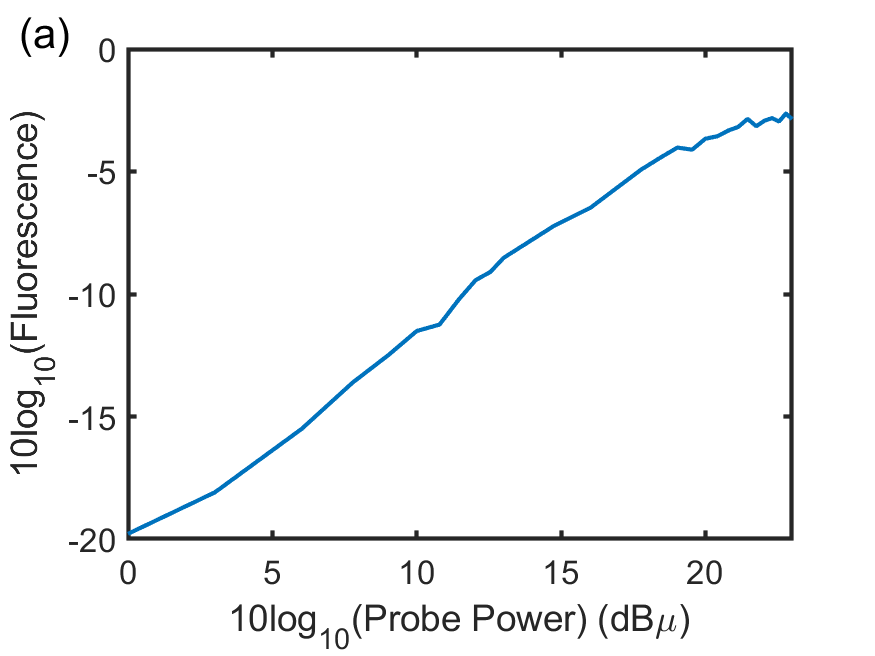}
    \caption{Fluorescence amplitude plotted against probe laser power in log scale in regime where electric field is weak (<1~mV/m).}
    \label{fig:resid}
\end{figure}

In the absence of an RF field, we observe a fluorescence response with increasing probe laser power, shown by Fig.~\ref{fig:resid}. This data is extracted from Fig.~\ref{fig:FLO_waterfall} (a) for the weakest RF field strength. As mentioned before, we measured the response with and without the RF power of -60 dBm and saw no change. This result allows us to reduce the expected sources of the residual peak. As we increase the probe laser power, we expect the number of Rydberg atoms to increase proportionally. In response to the increased number of atoms, the BBR-induced state transfer scales linearly with number of Rydberg atoms while the Rydberg-Rydberg collisions scale quadratically with number of Rydberg atoms. For this reason, we believe the source of this residual fluorescence is likely from BBR.

While the residual fluorescence is likely from BBR, we also consider is the collision of Rydberg atoms with ground state atoms (grd-ryd). This rate would also scale linearly with the number of Rydberg atoms. Isolating this will be a bit harder, but can be accounted for by analyzing the decay rate of the fluorescence, which is roughly 12 $\mu s$. This decay rate is a result of both BBR induced state transfer and grd-ryd collisions. We commented that the ryd-ryd collisions are on the order of 10*2$\pi$ kHz and the BBR induced state transfer is of the same order. This lifetime expected is roughly 50 $\mu s$ which corresponds to a decay rate of 20 kHz. Based on our measurement of 12 $\mu s$ fall time, we can find the remaining decay resulting from grd-ryd collisions is then
\begin{equation}
    \gamma_{col} = 1/\tau_{meas}-1/T_{BBR}-1/T_{ryd-ryd} \approx 60 kHz.
\end{equation}

While the residual fluorescence limits the sensitivity of the measurement, it allows for the determination of temperature. In future implementations, we will utilize the fluorescence to measure the ambient temperature. In this manuscript, we pulsed the RF field that transferred atoms from the 35P Rydberg state to the 34D Rydberg state. In the future, we will pulse the optical fields so that we can observe a rise time based on the population transfer induced by the BBR radiation. This may allow for a more precise measurement of the rise fall time.
\vspace{-5mm}

\section{Conclusion}
We have demonstrated measurements of Rydberg states using EIA/EIT and fluorescence. We found that the fluorescence measurement allowed for a more sensitive measurement by a factor of 4 in field. Another interesting feature was the bandwidth response of the three-photon EIA/EIT. We were able to show 3 MHz in bandwidth in this system. Finally, we worked to characterize a residual fluorescence present in the absence of the RF field. We found that this was the result of a combination of Rydberg atoms and ground state atoms and black body radiation. We find that this measurement may have potential for using the Rydberg atoms for thermometry (the measurements of temperature).

\begin{acknowledgments}
\noindent This work was partially funded by the NIST-on-a-Chip (NOAC) Program and was developed with funding from the Defense Advanced Research Projects Agency (DARPA).
 The views, opinions and/or findings expressed are those of the authors and should not be interpreted as representing the official views or policies of the Department of Defense or the U.S. Government. A contribution of the US government, this work is not subject to copyright in the United States.
\end{acknowledgments}

\section*{AUTHOR DECLARATIONS}
\vspace{-3mm}
{\bf Conflict of Interest}\\
The authors have no conflicts to disclose.
\vspace{10mm}

\section*{Data Availability Statement}
Data is available at https://datapub.nist.gov/od/id/mds2-3148.
Produces the bibliography via BibTeX.


\begin{thebibliography}{43}
\bibitem [{NIS()}]{NISTDisclaimer}%
  \BibitemOpen
  \href@noop {} {}\bibinfo {note} {Certain equipment, instruments, software, or materials are identified in this paper in order to specify the experimental procedure adequately. Such identification is not intended to imply recommendation or endorsement of any product or service by NIST, nor is it intended to imply that the materials or equipment identified are necessarily the best available for the purpose.}\BibitemShut {Stop}%
\end{thebibliography}
\end{document}